\documentclass[twocolumn]{openjournal}

\usepackage{xcolor}
\usepackage{textgreek}
\usepackage[utf8]{inputenc}
\usepackage[english]{babel}

\usepackage{hyperref}
\hypersetup{
    unicode, 
    colorlinks=true,
    linkcolor=linkcolor,
    citecolor=linkcolor,
    filecolor=linkcolor,
    urlcolor=linkcolor,
}
\usepackage{color,colortbl}
\definecolor{linkcolor}{rgb}{0.0,0.3,0.5}
\usepackage{tensind}
\tensordelimiter{?}
\DeclareGraphicsExtensions{.bmp,.png,.jpg,.pdf}
\usepackage{verbatim}
\usepackage[normalem]{ulem}
\usepackage{orcidlink}
\usepackage{soul}

\graphicspath{ {./figures/} }

\newcommand\Mgii{Mg\protect{\footnotesize{II}} }

\begin{document}
\title{The Giant Arc -- filament or figment?
\vspace{-1cm}}

\author{Till Sawala$^1$\orcidlink{0000-0003-2403-5358}}
\email[$^1$]{till.sawala@helsinki.fi}

\affiliation{Department of Physics, University of Helsinki, Gustaf H\"allstr\"omin katu 2, FI-00014 Helsinki, Finland}
\affiliation{Institute for Computational Cosmology, Durham University, South Road, Durham DH1 3LE, United Kingdom}

\author{Meri Teeriaho$^2$ \orcidlink{0000-0003-0070-7746}}
\email[$^2$]{meri.teeriaho@helsinki.fi}

\affiliation{Department of Physics, University of Helsinki, Gustaf H\"allstr\"omin katu 2, FI-00014 Helsinki, Finland}

\begin{abstract}
The so-called “Giant Arc” is a sparse pattern of \Mgii absorbers spanning $\sim$740 comoving Mpc, whose discovery has been claimed to contradict the large-scale homogeneity inherent to the standard cosmological model. We previously showed that, with the same algorithm and parameters used for its discovery, very similar patterns are abundant in uniform random distributions, and among equivalent halo samples in a cosmological simulation of the standard model. In a response, the original discoverers of the “Giant Arc” have argued that these parameters were only appropriate for their specific observational data, but that a smaller linking length should be used for control studies, in which case far fewer patterns are detected. We briefly review and disprove these arguments, and demonstrate that large patterns like the ``Giant Arc" are indeed ubiquitous in a statistically homogeneous universe.  
\vspace{.3cm}
\end{abstract}

\maketitle

\section{Introduction}\label{sec:intro}
The ``Giant Arc" was identified by \cite{Lopez-2022} as a three-dimensional FoF \citep[Friends-of-Friends, ][]{Huchra-1982, Davis-1985} group among a slice of \Mgii absorbers at redshift $z \sim 0.8$. \Mgii absorbers are a (very sparse) subset of galaxies detected due to their random alignment with background quasars \citep[e.g.][]{Chen-2008, Lundgren-2011, Gauthier-2014}. It has been interpreted as evidence of an anomaly or defect in the standard cosmological model \citep[e.g.][]{Aluri-2023, Lopez-2024b} or a signature of new physics \citep[e.g.][]{Lapi-2023, Constantin-2023, Mazurenko-2025, Meissner-2025}.

A crucial parameter of any FoF detection is the linking length, or more specifically, its relation to the typical interparticle distance. Lopez et al. gave no clear reason for their choice of 95~$c$Mpc (comoving Mpc) in their original work, except that after experimentation, it resulted in the clearest identification of the “Giant Arc”, a pattern that the same authors had previously noticed by eye.

In \cite{Sawala-2025}, we examined the claim that the ``Giant Arc" contradicts the large-scale homogeneity inherent to the standard model by searching for similar patterns both in {\sc Flamingo-10K}, a large cosmological simulation \citep{Schaye-2023, Pizzati-2024} in the $\Lambda$~Cold Dark Matter (hereafter $\Lambda$CDM) context, and in equivalent random samples created by a uniform Poisson process. Using the same value for the linking length and for several other parameters as \cite{Lopez-2022} had used, we found similar patterns ubiquitous both in the simulation and in the random samples. We also showed that the convex-hull overdensity that \citeauthor{Lopez-2022} had measured for the ``Giant Arc", and to which they had ascribed a $3.4 \sigma$ significance, is both typical of similar point patterns and unrelated to the underlying matter density. \cite{Lopez-2025} have responded to our work and claim that, due to several supposed shortcomings, we had made the ``Giant Arc" appear ordinary, when it is, according to them, truly exceptional.

It is not necessary to delve into speculations by \citeauthor{Lopez-2025} about possible insufficiencies of the {\sc Flamingo-10K} simulation. Suffice to say that, as a collisionless simulation, none of the mentioned numerical issues, which could potentially afflict hydrodynamic simulations, apply. As we had shown, for halo samples of this scarcity and scale, the simulation predicts a nearly uniform spatial distribution, and large patterns are almost equally abundant in random samples and in $\Lambda$CDM. For sake of argument and simplicity, in this work, we will consider only random samples, where any possible reservations about the simulation are irrelevant.

\section{Arguments for different linking lengths}
In their new work, \citeauthor{Lopez-2025} provide an alternative explanation for their original choice of a linking length of 95~$c$Mpc. In doing so, they simultaneously argue that, for a fair comparison, a linking length of only 65~$c$Mpc should be used for simulation data or uniform random samples. As expected, and as we had already shown in \cite{Sawala-2025}, they find that this reduced linking length in the controls yields far fewer large patterns, which would make the ``Giant Arc" seem exceptional again. Below, we briefly reproduce their chain of reasoning and point out several mistakes, before we explicitly test and disprove the implications.

The supposed starting point for the linking length is the mean nearest neighbour distance in uniform random samples, which turns out to be 65~$c$Mpc for the particular point density. It is not clear why this would be a good choice of linking length, and no explanation is given. \citeauthor{Lopez-2025} then argue that this linking length should be increased to 95~$c$Mpc to account for two effects present only in the observational data, but not when performing the control experiments with synthetic data.

\subsection*{Background Inhomogeneities}
The first proposed correction is meant to account for the inhomogeneous quasar background. To quantify the inhomogeneity, \citeauthor{Lopez-2025} calculate the average quasar count, $\mu = 15.6$, and its standard deviation, $\sigma = 5.7$, in squares of area $65^2$~$c$Mpc$^2$. Next, they declare that they ``reasonably consider" cells with a count of $1.5 \sigma$ below the mean to be underdense, and compute the ratio $\frac{\mu}{\mu  - 1.5 \sigma} = 2.23$ -- it is worth noting that this number is highly sensitive to the completely arbitrary definition of ``underdense". Without further explanation, they then increase the linking length by a factor of $2.23^{1/3} = 1.31$ from $65$~$c$Mpc to $85$~$c$Mpc to compensate for the inhomogeneity of the background, but argue that the smaller linking length of $65$~$c$Mpc should have been used in our random controls, which assumed a uniform background.

This correction and this distinction are not warranted. Let us first recall that directly selecting the sky positions of $k$ \Mgii absorbers from a uniform random distribution, as we had done for our control in \cite{Sawala-2025}, or selecting them as a subset of $n\geq k$ quasar positions previously drawn from the same random distribution, are identical. However, even a quasar background with no intrinsic inhomogeneities with the same mean number per cell has an expected standard deviation of $\sigma = \sqrt{\mu} = 3.95$, similar to the one \cite{Lopez-2025} measured in the real background. By the same reasoning, it would thus require a similar ``correction". Furthermore, additional inhomogeneities in the background do not reduce the chance of finding patterns in a random foreground, for which a larger linking length might compensate. To the contrary, and as we will show below, the rate of false-positives increases with increasingly inhomogeneous backgrounds.

\subsection*{Distance errors}
Secondly, \citeauthor{Lopez-2025} claim that the effect of redshift errors or peculiar velocities, which both manifest as distance errors in the foreground, should further increase the linking length by $\sim 10$~$c$Mpc. Without going into details, while it is true that distance errors would smear out a real structure and potentially necessitate a larger linking length to detect it, this also increases the incidence of false positives. As we will show, no distinction should be made between the data and the uniform random samples we use as a control.

Consider a uniform PDF $f(x) = \frac{1}{A}$ for $x \in [0,A]$. Suppose $e$ is a random error with PDF $f_E(e)$ and support in $[-B,B]$, and $B < A/2$, so that an interval $[B, A-B]$ can be defined. Then, if $z = x + e$ is a new random variable that includes the error, its PDF $f_z(z)$ over the interval $[B, A-B]$ is given by the convolution
$$
f_z(z) = \int_{-\infty}^{\infty} f_x(z-e) f_E(e) \ d e  = \frac{1}{A} \int_{-B}^{B}  f_E(e) \ d e = \frac{1}{A}
$$ where the second equality is due to the compact support of $f_x$, and the third equality is due to the compact support of $f_E$. In other words, the probability density $f_z(z)$ is identical to the original $f(x)$, independent of the shape of $f_E$ (e.g. uniform, Gaussian, etc.). In particular, if random errors with $B \ll A$ are applied to a uniform distribution, except from small boundaries of width $B$, the initial distribution is maintained.

The distance errors invoked by \cite{Lopez-2025} are $5-10$~$c$Mpc. If we assume that the uniformity which we had already assumed for a slice of thickness $338$~$c$Mpc extends by at least this small amount on both sides, applying random errors has exactly zero effect on the PDF, but even without this extension, the effect would be minuscule. In other words, our uniform random samples already account for arbitrary random errors. Increasing the linking length to account for errors in the observations, but using a smaller value for the random control, is incorrect.

According to \cite{Lopez-2025}, the combined result of both claimed effects is a linking length of 95~$c$Mpc, i.e. precisely the value on which 
\cite{Lopez-2022} had landed in their earlier paper via a completely different route. At this point, we could also note that the background density could only affect the distribution on the sky (and hence only two components of the spatial distribution), while distance errors could only affect the remaining, line-of-sight dimension. Yet, to arrive at a three-dimensional linking-length of 95~$c$Mpc, the two effects were simply added together. 

After explaining why different linking lengths should be used,
\citeauthor{Lopez-2025} proceed to use our public code and the data we provided to replicate our control using the much smaller linking length of 65~$c$Mpc. Unsurprisingly, they find that structures like the ``Giant Arc", which they had found using a value of 95~$c$Mpc, would then be exceedingly rare.

\begin{figure*}

\includegraphics[width=6.1cm, trim={.2cm 1.9cm 1.9cm 1.4cm},clip]{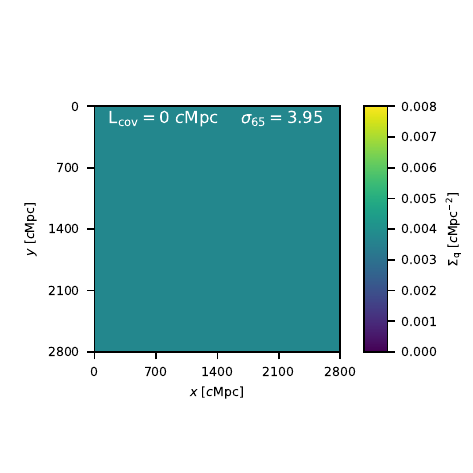}
\includegraphics[width=4.86cm, trim={1.4cm 1.9cm 1.9cm 1.4cm},clip]{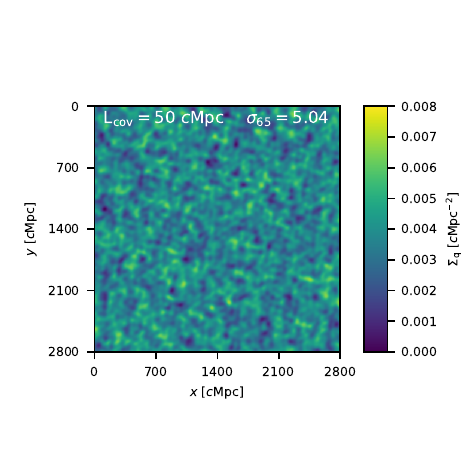}
\includegraphics[width=6.73cm, trim={1.4cm 1.9cm 0.1cm 1.4cm},clip]{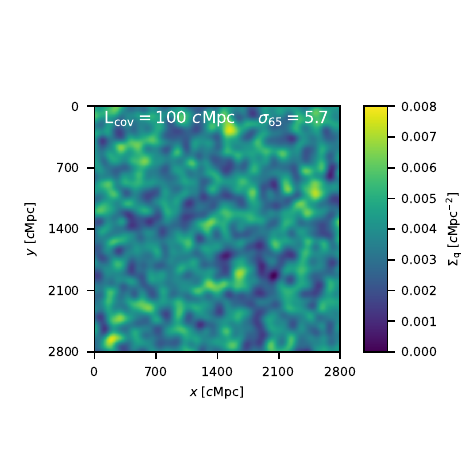} \\

\includegraphics[width=6.1cm, trim={.2cm 1.2cm 1.9cm 1.4cm},clip]{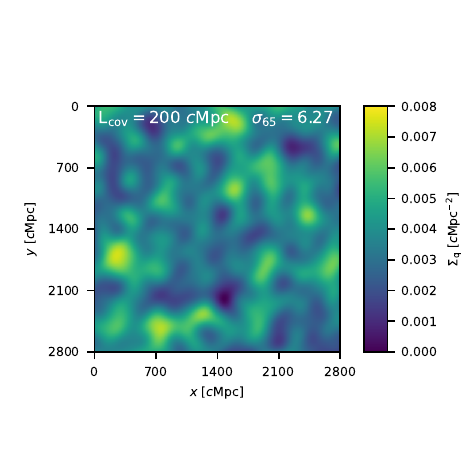}
\includegraphics[width=4.86cm, trim={1.4cm 1.2cm 1.9cm 1.4cm},clip]{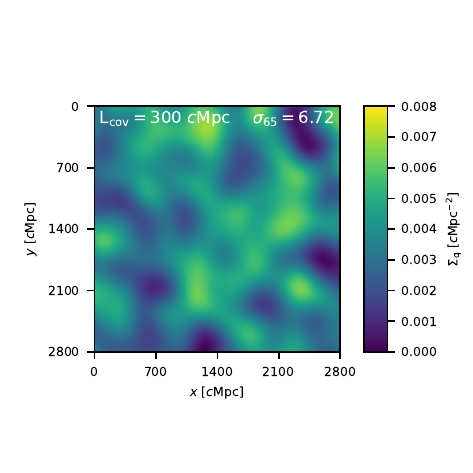}
\includegraphics[width=6.73cm, trim={1.4cm 1.2cm 0.1cm 1.4cm},clip]{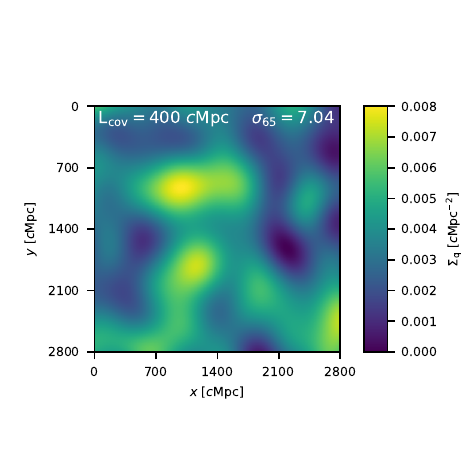}

\caption{\label{fig:i} Realisations of random fields of covariance lengths, $\mathrm{L_{cov}}$, ranging from 0 (homogeneous) to 400~$c$Mpc, representing increasingly inhomogeneous background densities, coloured here in terms of quasar surface density, $\Sigma_q$. $\sigma_{65}$ denotes the standard deviation in quasar number counts in cells of area 65$^2$~$c$Mpc$^2$ averaged over 1000 realisations.}
\end{figure*}

\section{The actual effect of background inhomogeneity and errors}
Fortunately, there is no need to further examine the arguments given for using different linking lengths for different data, as the implications can be tested directly. \citeauthor{Lopez-2025} effectively contend that large patterns, which we found so frequently in random samples, are {\it more} likely to appear in front of the uniform random background we had assumed, and {\it less} likely in front of the inhomogeneous quasar background of the real universe. This would make the ``Giant Arc" much more exceptional than we had concluded. However, as we will now demonstrate, this is incorrect. In fact, an inhomogeneous background only {\it increases} the chance of finding patterns in the foreground.

We start by creating backgrounds as realisations of Gaussian random fields with constant variance and variable covariance lengths, $\mathrm{L_{cov}}$. For consistency with \cite{Lopez-2025}, our backgrounds have the same mean number density, $\Sigma_\mathrm{q} = 0.0037~c$Mpc$^{-2}$, and we measure the standard deviation of quasar counts per cell of area $65^2$~$c$Mpc$^{2}$, which we denote as $\sigma_{65}$. In Fig.~\ref{fig:i}, we show six individual realisations with $\mathrm{L_{cov}}$ between 0 (i.e. homogeneous) and 400~$c$Mpc, resulting in average values of $\sigma_{65}$ between $3.95$ ($\mathrm{L_{cov}} = 0$) and $7.04$ ($\mathrm{L_{cov}} = 400~c$Mpc). The value of $\sigma_{65} = 5.7$ measured by \citeauthor{Lopez-2025} is found at $\mathrm{L_{cov}} = 100~c$Mpc.

We create 1000 independent realisations of the background at each covariance length, and use them as the intensity function for creating new foreground samples representing \Mgii absorbers in a weighted Poisson process. For each background, we generate 100 independent foregrounds, and choose the line-of-sight positions from uniform random samples. For completeness, we also add random errors, uniform over the interval $[-10, 10]$~$c$Mpc, to the line-of-sight positions, with absolutely no effect on the expected outcome. As in \cite{Sawala-2025}, we consider slices of $2800 \times 2800 \times 338 c$Mpc$^3$ with the same number density and thickness as in \cite{Lopez-2022}, and detect patterns as FoF-groups using a constant linking length of 95~$c$Mpc.

\begin{figure*}
\centering 
\includegraphics[width=17.cm, trim={0.0cm 0.0cm 0.0cm 0.0cm},clip]{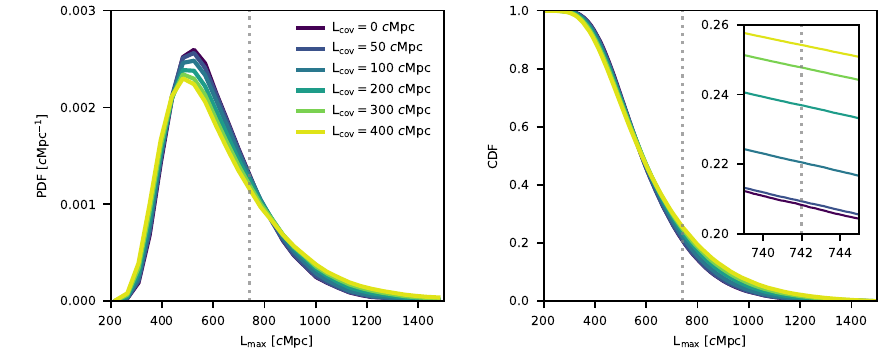}

\caption{\label{fig:ii} Probability density (left) and cumulative distribution (right) for the length ,$\mathrm{L_{max}}$, of the longest anisotropic ($b/a < 0.4$) pattern per slice of $2800\times2800\times338$~$c$Mpc$^3$ under different assumptions for the inhomogeneity of the background density. Dotted vertical lines denote 742~$c$Mpc, the extent of the ``Giant Arc". Foreground patterns as long or longer than the ``Giant Arc" are common for all backgrounds, and slightly {\it more} common for more inhomogeneous backgrounds. }

\end{figure*}

In Fig.~\ref{fig:ii}, we show the measured distribution of the length of the longest anisotropic pattern found in each foreground sample. For the anisotropy, we choose here $b/a < 0.4$ and refer the reader to \cite{Sawala-2025} for the definition and a visual comparison of similar filamentary patterns to the ``Giant Arc". We note that the actual size of the ``Giant Arc" was only vaguely described as ``approximately 1~Gpc" in \cite{Lopez-2022}. Because the authors had not been forthcoming with their data, in \cite{Sawala-2025} we could only estimate its extent, which we defined as the maximum pairwise separation between members, to be approximately 800~$c$Mpc. This turns out to be a slight overestimate, as \cite{Lopez-2025} now quote a value of only 742~$c$Mpc. When we adopt this lower value, the incidence of anisotropic patterns more extended than the ``Giant Arc" increases, to $\sim 20\%$ of samples with a homogeneous background.

If the arguments for changing linking lengths given in \citeauthor{Lopez-2025} were true, we should find far {\it fewer} large patterns with inhomogeneous backgrounds. As can be seen in Fig.~\ref{fig:ii}, the opposite is true. In fact, our assumption of a homogeneous background was conservative, and systematically {\it more} patterns are found in front of increasingly inhomogeneous backgrounds.

\section{Conclusion}
We have demonstrated the arguments for choosing drastically different linking lengths for the discovery of the ``Giant Arc" and for control studies, put forth in \cite{Lopez-2025} to be incorrect. Patterns like the ``Giant Arc" are common in uniform random distributions, and become even more common when background inhomogeneities are included. As we had already shown in \cite{Sawala-2025}, on scales beyond a few hundred Mpc, the spatial distribution of sparse halo samples in a $\Lambda$CDM universe is very close to uniform random, so our results directly apply there, too.

We note that we cannot test all possible backgrounds. However, after a uniform background, assuming statistical homogeneity and isotropy is the next most conservative assumption. Making quasar background anisotropic would make anisotropic patterns even more frequent, and also directly contradict observations \citep{Goyal-2024}. Moreover, were the ``Giant Arc" only exceptional in front of a very specific background, it also would not be a real foreground structure. And just to be very clear, the effect of mild background inhomogeneities on the already high chance of finding large patterns among samples of \Mgii absorbers, shown in Figure~\ref{fig:ii}, is small, so the appearance of the ``Giant Arc" must {\it not} now be taken as evidence of large-scale correlations in the quasar background.

The new justification of the choice of linking length presented in \cite{Lopez-2025} is incorrect. By contrast, the reason given in their original work, that it simply maximised the discovery in their particular data, appears more reasonable. However, as we had already demonstrated in \cite{Sawala-2025}, the same linking length that maximised the discovery of the ``Giant Arc" is also ideal for finding spurious patterns in noise. Here we show that this holds also for inhomogeneous backgrounds, where such patterns arise even more frequently. This affirms our conclusion that the ``Giant Arc" is almost certainly not a real feature of the underlying galaxy distribution. It is entirely consistent with being an apophenic pattern seen in an extremely sparse sample of the statistically homogeneous large-scale galaxy distribution predicted by the standard model.

Our work does not directly address the ``Big Ring", a second large pattern of \Mgii absorbers announced in \cite{Lopez-2024}. However, at least its FoF-based detection uses the same heuristics as was used for the ``Giant Arc", and the same concerns about both the choice of linking length and the potential for false positives apply.

Of course, our work does not show that the universe is homogeneous. It does demonstrate that detections of apparent inhomogeneities should be rigorously scrutinised. Adjusting parameters to confirm prior hypotheses already carries all the risks associated with the look-elsewhere effect. Subsequently using different parameters for the crucial null-hypothesis test is a dangerous path that can lead to false conclusions. In these circumstances, simply declaring a high statistical significance is insufficient. Controls of the kind we provide here \citep[see also][]{Nadathur-2013} should always accompany similar claims. \\
\  \\

\section*{Data Availability Statement}
Code to produce the results and figures presented here is provided at: \url{https://github.com/TillSawala/FilamentOrFigment}.

\section*{Acknowledgements}
We thank Carlos Frenk, Adrian Jenkins, Peter Johansson, Seshadri Nadathur, Gabor Racz, Alexander Rawlings and Joop Schaye for very helpful discussions and comments, and we are grateful to the anonymous reviewer for their further very helpful suggestions. T.S. and M.T. acknowledge support by the Research Council of Finland grants 354905 and 339127. T.S. acknowledges support by the European Research Council (ERC) through Consolidator Grant KETJU (no. 818930) and the Advanced Investigator grant DMIDAS (GA 786910), and support by the STFC Consolidated Grant ST/T000244/1. This work used facilities hosted by the CSC -- IT Centre for Science, Finland.

We gratefully acknowledge the use of open source software, including \texttt{Matplotlib} \citep{matplotlib-paper}, \texttt{SciPy} \citep{SciPy}, \texttt{Scikit-learn} \citep{scikit-learn}, \texttt{NumPy} \citep{numpy-paper} and \texttt{spam} \citep{SPAM}. 
\newpage

\bibliographystyle{apsrev4-1}
\bibliography{bibliography}

\  \\
\vspace{8cm} 
\ 

\end{document}